\documentclass[12pt]{article}
\usepackage{amssymb}
\begin{document}
\newfont{\elevenmib}{cmmib10 scaled\magstep1}%
\newfont{\cmssbx}{cmssbx10 scaled\magstep3}
\newcommand{\preprint}{
            \begin{flushleft} 
			\elevenmib Yukawa\, Institute\, Kyoto\\
            \elevenmib  University\, of\,  Durham
            \end{flushleft}\vspace{-1.3cm}
            \begin{flushright}\normalsize  \sf
            YITP-98-29\\ 
			DTP-98-27\\
			{\tt hep-th/9805106} \\ May 1998
            \end{flushright}}
\newcommand{\Title}[1]{{\baselineskip=26pt \begin{center} 
            \Large   \bf #1 \\ \ \\ \end{center}}}
\newcommand{\Author}{\begin{center}\large \bf
            A.\, J.\, Bordner$^a$, E.\, Corrigan$^b$\ \
            and R.\, Sasaki$^a$\end{center}}
\newcommand{\Address}{\begin{center} \it 
            $^a$ Yukawa Institute for Theoretical Physics, Kyoto
            University,\\ Kyoto 606-8502, Japan \\
            $^b$ Department of Mathematical Sciences, University of 
            Durham,\\ South Road, Durham DH1-3LE, United Kingdom
		    \end{center}}
\newcommand{\Accepted}[1]{\begin{center}{\large \sf #1}\\
            \vspace{1mm}{\small \sf Accepted for Publication}
            \end{center}}
\baselineskip=20pt

\preprint
\thispagestyle{empty}
\bigskip
\bigskip
\bigskip
\Title{Calogero-Moser Models: A New Formulation}
\Author

\Address
\vspace{2cm}

\begin{abstract}%
\noindent
A new formulation of Calogero-Moser models based on  root systems
and their Weyl group is presented. 
The general construction  of the Lax pairs 
applicable to all  models based 
on the simply-laced algebras (ADE)   
are given for two types which we  call 
`root'  and  `minimal'.
The root type  Lax pair is new; the matrices used in its construction 
bear a resemblance to the adjoint 
representation of the associated Lie algebra, and exist for all models,
but they do not contain elements associated with the zero weights 
corresponding to the Cartan subalgebra.
The root type provides a  simple method of constructing
sufficiently many number of conserved quantities  
for all models, including the one based on $E_{8}$, whose 
integrability had been an unsolved problem for more than twenty years.
The minimal types provide a unified description of 
all  known examples of Calogero-Moser Lax pairs and add some more.
In both cases, the root type and the minimal type,
the formulation works for all of the four choices of  
potentials: the rational, trigonometric, hyperbolic and elliptic.
\\ \\

\end{abstract}

\newpage

\section{Introduction}
\setcounter{equation}{0}
The Calogero-Moser models are one-dimensional 
dynamical systems with long-range pair-wise interactions.
They are completely integrable when the two-body interaction potential
\cite{CalMo,OP1,OP2,Ino}
is proportional to (i) $1/L^2$, (ii) $1/\sin^2L$, (ii) $1/\sinh^2L$
and (iv) $\wp(L)$, in which $L$ is the inter-particle ``distance''.
The types of integrable many-particle interactions are governed by
Lie algebras or rather their root systems:
there are Calogero-Moser models based on the root systems
of all of the semi-simple 
Lie algebras.  
The total number of the particles in the system is equal to the rank 
of the algebra and it is an arbitrary integer $r$ for
the classical Lie-algebras: $A_r$, $B_r$, $C_r$ and $D_r$. 
But it is quite limited for the exceptional algebras: 
$E_6$, $E_7$, $E_8$, 
 $F_4$ and $G_2$.

Since the models are  generic they find
various physical applications ranging from solid state physics to
particle physics \cite{ss}.
Because of their Lie algebraic structure, elliptic Calogero-Moser models
are  analysed intensively in connection with the 
Seiberg-Witten curve and 
differential for ${\cal N}=2$ supersymmetric gauge theory with the 
same Lie algebra \cite{SeiWit}-\cite{DHPh1}.

In this paper we address  the fundamental problems of the 
Calogero-Moser models rather than the applications.
These are the issues of integrability and the universal framework for
the construction of the Lax pairs.
These have been a mystery from the early days of the Calogero-Moser 
models. From the very beginning, 
the structure of the integrable Hamiltonians 
for all the models based on root systems of semi-simple Lie algebras
was understood but the catalogue of the Lax pairs necessary for the 
proof of the integrability remained the same for some twenty years.
It contained only the vector representations of the classical algebras
\(A_{r}\), \(B_{r}\), \(C_{r}\), \(D_{r}\) and the \(BC_{r}\) root 
system.
A general principle governing all the models based on the classical 
as well as the exceptional algebras was yet to be found.
Recently D'Hoker and Phong \cite{DHPh} succeeded in constructing Lax 
pairs for the exceptional algebras but the method was not 
complete enough to cover the one based on \(E_{8}\).

We present in this paper a new formulation of Calogero-Moser Lax pairs 
based on root systems and their Weyl group.
It is applicable to all models based on semi-simple Lie algebras 
including \(E_{8}\).
Constructions of Lax pairs
for models based on simply-laced root systems are given as two types
which we will call `minimal' and `root'. 
The root type of Lax pair is new; 
the matrices used in its construction bear a resemblance to the adjoint 
representation of the associated Lie algebra, and exist for all models,
but they do not contain elements associated with the zero weights 
corresponding to the Cartan subalgebra.
The `minimal' types provide a unified description of all known 
examples of Calogero-Moser Lax pairs and reveal some new ones.

This paper is organised as follows.
In section two the basic ingredients of the models are introduced and 
the principle of Weyl invariance is stated.
In section three the Lax pair of the `root' type
for  simply-laced root systems is given and its consistency 
is proved.
The Lax pairs of the models for the non-simply laced root systems can
be obtained by reduction or folding \cite{folding}, which is a well-known 
procedure in Toda lattice (field) theory \cite{Toda, TodaSmatrices}.
In section four the formulation of the Lax pairs of the minimal 
type is given and consistency is proved in a similar way 
as in the root type.
In both cases, the root type and the minimal type,
the Lax pair exists for all the four types of the interaction
potentials.
Various known examples of Lax pairs are derived as special cases of the
minimal type.
Some Lax pairs of non-simply laced root systems are derived from those of 
simply-laced ones by reduction or folding.
The spinor and the anti-spinor representations of \(D_{N}\) are 
discussed in some detail for two purposes;
the first  to exemplify the relationship between
the exponents of the algebra and conserved quantities and
the second to derive the $B_N$ Lax pair in the spinor representation 
by reduction.
The Lax pairs to be discussed in this paper are those without 
 spectral parameter.
Introduction of the spectral parameter to the elliptic 
potential case in the present scheme is rather straightforward
\footnote{For the  Lax-pair with a  spectral parameter the
see for example \cite{DHPh,Krie}.}.
We will discuss the Lax pair with spectral parameter in connection 
with folding in a future publication.
Section five is for summary and discussion.

\section{Calogero-Moser Models}
\setcounter{equation}{0}
Let us start by defining the Calogero-Moser model
based on a semi-simple and {\em simply-laced} Lie algebra 
${\mathfrak g}$ with rank $r$. 
In fact we only need the data of its roots.
We denote the set of all roots by $\Delta$. They are real $r$ 
dimensional vectors
and are normalised, without loss of generality, to 2:
\begin{equation}
	\Delta=\{\alpha,\beta,\gamma,\ldots\}, \quad \alpha\in 
{\mathbb R}^r,\quad
	\alpha^2=\alpha\cdot\alpha=2,\quad \forall\alpha\in\Delta.
	\label{eq:setroots}
\end{equation}
We denote by $Dim$ the total number of roots of $\Delta$. 
It is $r(r+1)$ and $2r(r-1)$ for $A_r$ and $D_r$ and 
72, 126 and 240 for $E_6$, $E_7$ and $E_8$, respectively.

The dynamical variables are canonical coordinates $\{q^j\}$ and their 
canonical conjugate momenta $\{p_j\}$ with the Poisson brackets:
\begin{equation}
	q^1,\ldots,q^r, \quad
p_{1},\ldots,p_{r}, \quad
\{q^j,p_{k}\}=\delta_{j,k},\quad \{q^j,q^{k}\}=\{p_{j},p_{k}\}=0.
	\label{eq:poisson}
\end{equation}
In most cases we denote them by $r$ dimensional vectors $q$ and $p$
\footnote{
For  $A_r$ models, it is customary to introduce one more degree of 
freedom,
$q^{r+1}$ and $p_{r+1}$ and embed 
all of the roots in ${\mathbb R}^{r+1}$.},
\begin{displaymath}
	q=(q^1,\ldots,q^r)\in {\mathbb R}^r,\quad
	p=(p_1,\ldots,p_r)\in {\mathbb R}^r,\quad
\end{displaymath}
so that the scalar products of $q$ and $p$ with the roots 
$\alpha\cdot q$, $p\cdot\beta$, etc. can be defined.

Another ingredient of the theory are the Weyl reflections.
Let $\xi$ be an ${\mathbb R}^r$ vector and $\beta\in \Delta$.
The Weyl reflection by a root $\beta$ is defined by
\begin{equation}
	W_{\beta}(\xi)=\xi-{2(\beta\cdot\xi)\beta\over{\beta^2}}.
	\label{eq:weyl}
\end{equation}
Obviously $W_{\beta}^2=1$ and $W_{\beta}=
W_{-\beta}=W_{\beta}^{-1}$
and the totality of the Weyl reflections 
form a group called the Weyl group.
The root systems of the semi-simple Lie algebras are invariant under  
any Weyl reflection: 
\begin{equation}
	W_{\beta}(\alpha)\in\Delta, \quad \forall 
	\alpha,\beta\in \Delta.
	\label{eq:weylinv}
\end{equation}
In fact, the set of roots invariant under the Weyl 
reflection (\ref{eq:weylinv})
is the fundamental ingredient for constructing a Calogero-Moser model.  
The root system need not belong to a Lie algebra.
The Lie algebra structure is important for 
most cases but not essential.
This can be seen most clearly in the $BC_r$ Calogero-Moser model, in 
which the set of roots $\Delta$ is the union of $B_r$ and $C_r$ roots.

Next we introduce the functions appearing in the Lax pair. 
They depend on the choice of the inter-particle potential.
For  the rational, $1/L^2$, potential they are:
\begin{equation}
	x(t)=x_{r}(t)={1\over t},\quad y(t)=y_{r}(t)=-{1\over{t^2}},
	\quad z(t)=z_{r}(t)=-{1\over{t^2}}.
	\label{eq:functions}
\end{equation}
For  the trigonometric, $1/\sin^2L$, potential they are:
\begin{equation}
		x(t)=x_{r}(t)=a\cot at,\quad y(t)=y_{r}(t)=-{a^{2}\over{\sin^2 at}},
			\quad z(t)=z_{r}(t)=-{a^{2}\over{\sin^2 at}},\quad a: const.
	\label{eq:functionstri}
\end{equation}	
For  the hyperbolic, $1/\sinh^2L$, potential they are:
\begin{equation}
		x(t)=x_{r}(t)=a\coth at,\quad y(t)=y_{r}(t)=-{a^{2}\over{\sinh^2 at}},
			\quad z(t)=z_{r}(t)=-{a^{2}\over{\sinh^2 at}}.
	\label{eq:functionshyp}
\end{equation}	
For  the elliptic, $ \wp(L)$, potential there are several choices of
the functions. Generally the functions $x$ and $x_{r}$ 
differ. A first choice is
\begin{eqnarray}
		x(t)&=&{a\over2}\left[ {{1+k\, {\rm sn}^2(at/2,k)}
		\over {\rm sn}(at/2,k)}-
i{{(1+k)(1-k\, {\rm sn}^2(at/2,k))}\over{{\rm cn}(at/2,k)\,
{\rm dn}(at/2,k)}}\right],\nonumber\\
y(t)&=&x^\prime(t),
			\quad z(t)=-b^2\wp(bt),\quad b=a/\sqrt{e_{1}-e_{3}},
	\label{eq:functionsell1}
\end{eqnarray}
and
\begin{equation}
		x_{r}(t)={a\over{{\rm sn}(at,k)}},\quad 
		y_{r}(t)=-a^{2}{{\rm cn}(at,k)\,{\rm dn}(at,k)\over{{\rm sn}^2(at,k)}},
			\quad z_{r}(t)=-b^2\wp(bt),\quad b=a/\sqrt{e_{1}-e_{3}},
	\label{eq:functionsell2}
\end{equation}
in which $k$ is the modulus of the elliptic function
\footnote{The detailed properties of the elliptic potential cases
 will be discussed elsewhere.}.\\
A second choice is
\begin{eqnarray}
		x(t)&=&{a\over2}\left[ {{{\rm cn}^{2}(at/2,k)-k^{\prime}{\rm sn}^2(at/2,k)}
		\over {\rm sn}(at/2,k)\,{\rm cn}(at/2,k)}+
(1+k^{\prime}){{\rm cn}^{2}(at/2,k)+k^{\prime}{\rm 
sn}^2(at/2,k)\over{{\rm dn}(at/2,k)}}\right],\nonumber\\
y(t)&=&x^\prime(t),
			\quad z(t)=-b^2\wp(bt),
	\label{eq:functionsell3}
\end{eqnarray}
and
\begin{equation}
		x_{r}(t)=a\,{{\rm cn}(at,k)\over{{\rm sn}(at,k)}},\quad 
		y_{r}(t)=-a^{2}{{\rm dn}(at,k)\over{{\rm sn}^2(at,k)}},
			\quad z_{r}(t)=-b^2\wp(bt),%
	\label{eq:functionsell4}
\end{equation}
in which $k^{\prime}=\sqrt{1-k^{2}}$.\\
A third choice is
\begin{eqnarray}
		x(t)&=&{a\over2}\left[ {{{\rm dn}^{2}(at/2,k)+ikk^{\prime}{\rm sn}^2(at/2,k)}
		\over {\rm sn}(at/2,k)\,{\rm dn}(at/2,k)}+
{k\,{\rm cn}^{2}(at/2,k)-ik^{\prime}\over{{\rm cn}(at/2,k)}}\right],\nonumber\\
y(t)&=&x^\prime(t),
			\quad z(t)=-b^2\wp(bt),
	\label{eq:functionsell5}
\end{eqnarray}
and
\begin{equation}
		x_{r}(t)=a\,{{\rm dn}(at,k)\over{{\rm sn}(at,k)}},\quad 
		y_{r}(t)=-a^{2}{{\rm cn}(at,k)\over{{\rm sn}^2(at,k)}},
			\quad z_{r}(t)=-b^2\wp(bt),%
	\label{eq:functionsell6}
\end{equation}
The trigonometric  ($k\to0$)  and hyperbolic ($k\to1$) limits of the
elliptic cases give other sets of functions for these cases. 
One important property is that they all satisfy the {\em sum rule}
\begin{equation}
	y(u)x(v)-y(v)x(u)=
	x(u+v)[z(u)-z(v)],
	\quad u,v\in {\mathbb C}.
	\label{eq:ident1}
\end{equation}
The functions $x_r$, $y_{r}$ and $z_r$ satisfy the same relations.
In all these cases the inter-particle potential \(V\) is proportional to
\(-z + const\) and $y$ ($y_{r}$) is the derivative of $x$ ($x_r$) and 
$z$ is always an even function:
\begin{equation}
	y(t)=x^\prime(t),\quad z(t)=x(t)x(-t)+constant,\quad
	 z(-t)=z(t).
	\label{eq:parity}
\end{equation}
For the rational (\ref{eq:functions}), trigonometric 
(\ref{eq:functionstri}) and hyperbolic cases (\ref{eq:functionshyp})
$x$ is an odd function and $y$ is 
an  even function 
but they do not have definite parity for the elliptic potentials 
(\ref{eq:functionsell1}), (\ref{eq:functionsell2}).

\bigskip

The Hamiltonian is given by ($g$ is a real coupling constant)
\begin{equation}
	{\cal H}={1\over2}p^2-{g^2\over2}\sum_{\alpha\in\Delta}
	x(\alpha\cdot q)x(-\alpha\cdot q),
	\label{eq:hamiltonian}
\end{equation}
which is invariant under the Weyl reflection of the dynamical 
variables:
\begin{equation}
	q\to q^\prime=W_{\beta}(q),\quad
	p\to p^\prime=W_{\beta}(p),\qquad
	\forall \beta\in\Delta,
	\label{eq:weylpq}
\end{equation}
forming a discrete subgroup of $O(r)$. 
In fact, ${p^\prime}^2=p^2$ and $x(\alpha\cdot 
q^\prime)=x(W_{\beta}(\alpha)\cdot q)$ and the invariance of $\Delta$ 
under Weyl group (\ref{eq:weylinv}) is used.
Let us compare the situation with the Toda lattice (field 
theory) \cite{Toda, TodaSmatrices}, another well-known integrable 
system based on the root systems of (affine) Lie algebras. 
In the latter, only 
the simple roots are used and  Weyl invariance does not exist
\footnote{Though Weyl invariance is absent, the Coxeter element
(a product of Weyl transformations corresponding to simple roots) 
plays an important role
\cite{Coxeter}}.
It should be stressed that in both cases, Calogero-Moser and Toda,
the Hamiltonian is not  invariant under the Lie algebra ${\mathfrak g}$ 
associated with the root system.
However, in Toda theories the Lax pairs and the classical R-matrices
are constructed from Lie algebra generators and therefore 
automatically work in any representation.
For Calogero-Moser models several attempts \cite{OP1,OP2,DHPh} to 
generate the Lax pairs based on Lie algebra generators and/or  
symmetric space ideas have not achieved the desired goal.

\bigskip
In this paper we propose to adopt the root systems and 
their Weyl invariance rather than the Lie algebraic structure 
as the basic principle of the Calogero-Moser models. 
Thus in order to find the Lax pair for the above Hamiltonian we should look 
for a space in which {\em the Weyl reflections rather than the Lie algebra 
generators} are conveniently represented.
Obviously the simplest and thus the  best choice is the set of roots 
$\Delta$ itself. The Lax pairs thus constructed  will be called 
of the `root' type.
It should be stressed that this is {\em different} from the adjoint 
representation.
The adjoint representation has $Dim + r$ dimensions. That is, it has
rank ($r$) number of zero weights corresponding to the Cartan 
subalgebra. These zero weights cause severe problems in representing 
the Weyl reflection in a $Dim + r$ dimensional linear space
when $r>1$, because the representation matrix can never be uniquely 
determined in the $r$ dimensional subspace.
This is the main obstacle for the proof of the integrability of
the $E_8$
theory,  for which the lowest dimensional Lie-algebra
representation is the  adjoint representation and as we will see  in
section four the minimal representation does not exist.

\section{Lax Pair of the `Root' Type}
\setcounter{equation}{0}

In this section we present the construction of the Lax pair 
applicable to all of the Calogero-Moser models based on semi-simple and 
simply-laced algebras. This provides the basic ingredients for
a unified proof of integrability of all
Calogero-Moser models based on root systems of semi-simple Lie algebras, 
including those based on non-simply laced root systems.
The Lax pairs of non-simply laced theories are obtained from the
corresponding simply-laced ones by {\em reduction} or {\em folding},
a well-known procedure in Toda lattice (field) theory 
\cite{folding, TodaSmatrices}. The non-simply laced Lax pairs obtained by 
reduction of the simply laced ones have only one coupling constant.
The direct formulation of the root type Lax pairs  
for non-simply laced theories 
and the $BC_{r}$ root system  (with two or more independent coupling
constants) could be 
given in a similar way as in this paper. 
 
The goal is to express the canonical equation of motion derived from 
the  Hamiltonian
(\ref{eq:hamiltonian}) in an equivalent matrix form :
\begin{equation}
	\dot{L}={d\over{dt}}L=[L,M],
	\label{eq:laxeq0}
\end{equation}
so that a sufficient number of conserved quantities could be obtained
by the trace:
\begin{equation}
	{d\over{dt}}Tr(L^k)=0,\quad k=1,\ldots,.
\end{equation}
It should be noted that the Lax pair in all theories and in all 
representations has the gauge freedom (similarity 
transformation):
\begin{eqnarray}
	L\to L^U & = & U^{-1}LU,\quad M\to M^U= U^{-1}MU+U^{-1}\dot{U},
	\nonumber  \\
		\dot{L}&=&[L,M],\quad \Longleftrightarrow \quad  
		\dot{L^U}=[L^U,M^U]. 
	\label{eq:gauge}  
\end{eqnarray}
The following Lax pair in $\Delta$ is believed to be in the simplest
gauge  (we choose \(L\) to 
be hermitian and \(M\) anti-hermitian):
\begin{eqnarray}
	L(q,p) & = & p\cdot H + X + X_{r},\nonumber\\
	M(q) & = & D+Y+Y_{r}.
	\label{eq:genLaxform}
\end{eqnarray}
Here $L$, $H$, $X$, $X_{r}$, $D$, $Y$ and $Y_{r}$ are $Dim\times 
Dim$ matrices whose indices are labelled by the roots themselves, 
usually denoted by 
$\alpha$,
$\beta$, $\gamma$, $\eta$ and $\kappa$. 
$H$ and $D$ are diagonal:
\begin{equation}
	H_{\beta \gamma}=\beta \delta_{\beta, \gamma},\quad
	D_{\beta \gamma}= \delta_{\beta, \gamma}D_{\beta},\quad
	D_{\beta}=-ig\left(z(\beta\cdot q)+\sum_{\kappa\in\Delta,\  
	\kappa\cdot\beta=1}z(\kappa\cdot q)\right).
	\label{eq:HD}
\end{equation}
$X$ and $Y$ have the same form, differing only by the dependence on 
the coordinates $q$:
\begin{equation}
	X=ig\sum_{\alpha\in\Delta}x(\alpha\cdot q)E(\alpha),\quad
	Y=ig\sum_{\alpha\in\Delta}y(\alpha\cdot q)E(\alpha),\quad
	E(\alpha)_{\beta \gamma}=\delta_{\beta-\gamma,\alpha}.
	\label{eq:XYdef}
\end{equation}
$X_r$ and $Y_r$ are necessary only in the `root' type Lax pair
\begin{equation}
	X_r=2ig\sum_{\alpha\in\Delta}
x_{r}(\alpha\cdot q)E_{d}(\alpha),\quad
	Y_r=ig\sum_{\alpha\in\Delta}
y_{r}(\alpha\cdot q)E_{d}(\alpha),\quad
	E_{d}(\alpha)_{\beta \gamma}=\delta_{\beta-\gamma,2\alpha}.
	\label{eq:XYrdef}
\end{equation}
The functions $x,y,z$ ($x_r,y_r,z_r$) are listed in 
(\ref{eq:functions})--(\ref{eq:functionsell6}).
The matrix $E(\alpha)$ ($E_{d}(\alpha)$) might be called a (double)
root discriminator. It takes the value one only when the difference
of the two indices is equal to  (twice) the root $\alpha$.
Though the matrices \(H\) and \(E(\alpha)\) satisfy  relations
\begin{eqnarray}
	[H,E(\alpha)]&=&\alpha E(\alpha),\quad
\ [H,[E(\alpha),E(\beta)]]=(\alpha+\beta)[E(\alpha),E(\beta)],
\nonumber\\
\ E(-\alpha)&=&E(\alpha)^T,\quad 
[E(\alpha),E(-\alpha)]+2[E_d(\alpha),E_d(-\alpha)]=\alpha\cdot H,
	\label{eq:algebra}
\end{eqnarray}
they are {\em not Lie algebra generators}.
The matrix elements $X_{\beta\gamma}$ and $Y_{\beta\gamma}$ are 
non-vanishing only when $\beta-\gamma$ is a root.
For simply-laced root systems with (length)${}^2=2$, this can be 
rephrased as
 \begin{equation}
	 X_{\beta\gamma}=0 \quad 
\mbox{and} \quad  Y_{\beta\gamma}=0\quad 
	 \mbox{if}\quad \beta\cdot\gamma\neq1.
 	\label{eq:adjac}
 \end{equation}
It is easy to rewrite $D$ in a form similar to $X$ and $Y$,
\begin{equation}
	D=-ig\sum_{\alpha\in\Delta}z(\alpha\cdot q)K(\alpha),\quad
	K(\alpha)_{\beta \gamma}=\delta_{\beta, 
	\gamma}\left(\delta_{\alpha,\beta}+\theta
(\alpha\cdot\beta)\right) ,
	\label{eq:Kdef}
\end{equation}
in which $\theta(t)$ has a support only on 1

\begin{equation}
	\theta(t)=\left\{\begin{array}{ll}
				1, \quad & t=1,\\
	0,\quad & \mbox{otherwise}. 
	\end{array}\right.
	\label{theta}
\end{equation}

\bigskip
It is straightforward to represent the Weyl reflections in $\Delta$.
By {\boldmath $\alpha$}  we denote a $Dim$ 
dimensional vector whose elements 
are the roots themselves. 
The Weyl reflection
can be 
represented by a $Dim\times Dim$ matrix $S(\beta)$
as follows:
Under the  Weyl reflection in terms of a root 
$\beta$, $W_{\beta}$ (\ref{eq:weyl}), each root $\alpha$ is mapped to
$\alpha\to \alpha^\prime=W_{\beta}(\alpha)$.  
We express  the transformation of the totality of the roots
as
\begin{equation}
	\hbox{\boldmath $\alpha$}\to
	\hbox{\boldmath $\alpha$}^\prime=S(\beta)\,
\hbox{\boldmath $\alpha$}.
	\label{eq:SBdef}
\end{equation}
It is easy to see that the elements of $S(\beta)$ are expressed as
\begin{equation}
	S(\beta)_{\gamma \eta}=\delta_{\gamma, 
W_{\beta}(\eta)}, \quad 
	\forall \beta, \gamma,\eta\in\Delta.
	\label{eq:SBform}
\end{equation}
The matrices $E(\alpha)$ ($E_d(\alpha)$) and $K(\alpha)$ transform
\begin{equation}
	S(\beta)^{-1}E(\alpha)S(\beta)=E(W_{\beta}(\alpha)),\quad
	S(\beta)^{-1}K(\alpha)S(\beta)=K(W_{\beta}(\alpha)),\quad
	\forall\beta\in\Delta.
	\label{eq:weylinvXK}
\end{equation}
The Weyl covariance of $L$ and $M$ is a simple consequence of
(\ref{eq:weylinvXK}). That is for
\begin{eqnarray}
	q\to q^\prime=W_{\beta}(q),\quad
	& p\to p^\prime=W_{\beta}(p),\qquad
	\forall \beta\in\Delta,\nonumber\\
	L(q^\prime,p^\prime)=S(\beta)^{-1}L(q,p)S(\beta),\quad
	\mbox{and}\quad
	&M(q^\prime)=S(\beta)^{-1}M(q)S(\beta).
	\label{eq:LMcov}
\end{eqnarray}
Then the Lax equation 
\[
	\dot L={d\over{dt}}L=[L,M]
\]
is invariant.
\bigskip
It should be remarked that the covariance requirement on the matrix 
$E(\alpha)$ ($E_d(\alpha)$) and $K(\alpha)$ (\ref{eq:weylinvXK}) is 
very strong. Once $E(\alpha_0)$ ($E_d(\alpha_0)$) and $K(\alpha_0)$
is given for one root $\alpha_{0}$ all the other 
$E(\alpha)$ ($E_d(\alpha)$) and $K(\alpha)$ for all the roots $\alpha$ 
lying on the same Weyl orbit of $\alpha_{0}$
are determined uniquely. In the case of simply-laced root systems
there is only one Weyl orbit, so everything is determined.
Changing the signs of some of the matrices $E(\alpha)$ (except for 
the overall similarity transformation (\ref{eq:gauge}))
would destroy the Weyl covariance and thus the Lax pair.
This is in  contrast with the Lie algebra or symmetric space 
representations.
In these cases the choices of the generators are much less restricted.

\bigskip
In the rest of the section we show that the Lax equation
\begin{equation}
	\dot L={d\over{dt}}L=[L,M]
	\label{eq:laxeq}
\end{equation}
is equivalent to the canonical equations of motion for the Hamiltonian
(\ref{eq:hamiltonian}):
\begin{equation}
	\dot q=p,\quad \dot p=-{\partial\over{\partial q}}{\cal 
	H}=-{g^2\over2}\sum_{\alpha\in\Delta}
	\left(\phantom{\sum}\hspace{-0.5cm}x(\alpha\cdot 
	q)
y(-\alpha\cdot q)-x(-\alpha\cdot q)
y(\alpha\cdot q)\right)\alpha.
	\label{eq:caneq}
\end{equation}
The Lax equation (\ref{eq:laxeq}) is decomposed into three parts:
\begin{eqnarray}
	{d\over{dt}}(X+X_r) & = & [p\cdot H,Y+Y_r],
	\label{eq:qdot}  \\
	{dp\over{dt}}\cdot H & = &
[X+X_r,Y+Y_r]_{\mbox{\footnotesize  diagonal part}},
	\label{eq:pdot}  \\
	0 & = & [X+X_r,D+Y+Y_r]_{\mbox{\footnotesize
off-diagonal part}}.
	\label{eq:offd}
\end{eqnarray}
It is easy to see that (\ref{eq:qdot}) is equivalent to the first set 
of the canonical equations of motion
$\dot q=p$.
In fact, by taking $(\beta,\gamma)$ element of (\ref{eq:qdot}), we 
obtain
\begin{eqnarray}
	[p\cdot H,Y]_{\beta \gamma} & = & 
	ig\sum_{\alpha\in\Delta}y(\alpha\cdot q)
E(\alpha)_{\beta  \gamma}p\cdot(\beta-\gamma)
\nonumber  \\
	 & = & ig\sum_{\alpha\in\Delta}
y(\alpha\cdot q)E(\alpha)_{\beta 
	\gamma}p\cdot\alpha
	\nonumber \\
	 & = & {d\over{dt}}X_{\beta \gamma},
	\label{eq:qdotmat}
\end{eqnarray}
in which the relations $\dot q=p$ and $x^\prime=y$
are used. 
Similar relation holds for $X_r$.

By using (\ref{eq:adjac}),%
we arrive at
\begin{eqnarray*}
	[X,Y]_{\beta\beta} & = & \sum_{\kappa\in\Delta,\ \kappa\cdot\beta=1}
	(X_{\beta\kappa}Y_{\kappa\beta}-
Y_{\beta\kappa}X_{\kappa\beta}) \\
	 & = & -g^2\sum_{\kappa\in\Delta,\ \kappa\cdot\beta=1}\left[
	 x((\beta-\kappa)\cdot q)y((\kappa-\beta)\cdot q)-
	 y((\beta-\kappa)\cdot q)x((\kappa-\beta)\cdot q)\right]%
\end{eqnarray*}
Likewise we obtain
\begin{eqnarray*}
 [X,Y_r]_{\beta\beta} & = &[X_{r},Y]_{\beta\beta}=0, \\
 {[X_r,Y_r]}_{\beta\beta} & = &-2g^2\left(x_r(\beta\cdot q)y_r(-\beta\cdot q)-
x_r(-\beta\cdot q)y_r(\beta\cdot q)\right)\\
 & = &
-2g^2\left(x(\beta\cdot q)y(-\beta\cdot q)-
x(-\beta\cdot q)y(\beta\cdot q)\right).
\end{eqnarray*}
Thus  (\ref{eq:pdot}) reads
\begin{eqnarray}
\dot p\cdot\beta&=&-g^2
\left(\sum_{\kappa\in\Delta,\ \kappa\cdot\beta=1}
	x((\beta-\kappa)\cdot q)y((\kappa-\beta)\cdot q)
	-y((\beta-\kappa)\cdot q)x((\kappa-\beta)\cdot q)\right.\nonumber\\
	& & \left.\hspace{1cm}+
2x(\beta\cdot q)y(-\beta\cdot q)-2x(-\beta\cdot q)y(\beta\cdot 
q)\phantom{\sum_{\kappa\in\Delta,\ \kappa\cdot\beta=1}}\hspace{-1.5cm}
	\right)\nonumber\\
	&=&-g^2\left(\sum_{\alpha\in\Delta,\ \alpha\cdot\beta=1}
	x(\alpha\cdot q)y(-\alpha\cdot q)-
	x(-\alpha\cdot q)y(\alpha\cdot q)\right.\nonumber\\
	& &\left.\hspace{2cm}+2x(\beta\cdot q)y(-\beta\cdot q)-
	2x(-\beta\cdot q)y(\beta\cdot q)
	\phantom{\sum_{\alpha\in\Delta,\ \alpha\cdot\beta=1}}
	\hspace{-1.5cm}\right),
	\label{eq:LMpdot}
\end{eqnarray}
in which the dummy variable is changed from $\kappa$ to 
$\alpha=\beta-\kappa$.
This equation is obtained from the second set of canonical equations
of motion
(\ref{eq:caneq}) by multiplying $\beta$ on both sides
\[
\dot p\cdot\beta=-{g^2\over2}\sum_{\alpha\in\Delta}
\left(\phantom{\sum}\hspace{-0.5cm}x(\alpha\cdot q)
y(-\alpha\cdot q)-x(-\alpha\cdot q)
y(\alpha\cdot q)\right)\alpha\cdot\beta.
\]
Only those terms corresponding to $\alpha\cdot\beta=\pm2$, ie, 
$\alpha=\pm\beta$ and $\alpha\cdot\beta=\pm1$ contribute 
and we obtain
(\ref{eq:LMpdot}). This leaves us to show the vanishing of 
(\ref{eq:offd}), which we decompose into four cases:
(A)  $\beta\cdot\gamma=1$ case, (B) 
$\beta\cdot\gamma=0$ case, (C) $\beta\cdot\gamma=-1$ case
and (D) $\beta\cdot\gamma=-2$ case. Let us evaluate
(\ref{eq:offd}) in turn.

\bigskip
\subsection{Consistency of the Root Type Lax Pair}
\subsubsection{(A) \(\beta\cdot\gamma=1\) case}

Let us start with
\[
[X,D]_{\beta \gamma}=X_{\beta \gamma}(D_{\gamma}-D_{\beta}),
\]
in which
\[
D_{\gamma}-D_{\beta}=-ig\left(z(\gamma\cdot q)-z(\beta\cdot 
q)+\sum_{\kappa\cdot\gamma=1}z(\kappa\cdot q)-
\sum_{\kappa'\cdot\beta=1}z(\kappa'\cdot q)\right).
\]
First we simplify the above expression by removing all of the 
cancelling terms. The first summation (\(\kappa\cdot\gamma=1\)) is 
decomposed into four groups according to the value of 
\(\kappa\cdot\beta=\{2,1,0,-1\}\). 
The term \(\kappa\cdot\beta=-2\) does 
not exist, since it means \(\kappa=-\beta\) which is incompatible with
\(\beta\cdot\gamma=1\) and \(\kappa\cdot\gamma=1\).
The term  \(-z(\beta\cdot q)\) cancels the
 \(\kappa=\beta\) (\(\kappa\cdot\beta=2\)) term 
in the first summation.
Likewise, \(z(\gamma\cdot q)\) term 
cancels the \(\kappa'=\gamma\) term in the 
second summation. 
The second group \(\kappa\cdot\beta=1\) in the first summation  
can be dropped since it is canceled by the term in the 
second sum having \(\kappa'\cdot\beta=1\).
The fourth group \(\kappa\cdot\beta=-1\) 
with \(\kappa\cdot\gamma=1\) 
consists of one term, since this means \(\kappa\cdot(\gamma-\beta)=2\)
implying \(\kappa=\gamma-\beta\). Then the term 
\(z((\gamma-\beta)\cdot q)\) is cancelled by \(z((\beta-\gamma)\cdot 
q)\) term in the second summation (\(z\) is an even function).
Thus only the third group survives:
\[
D_{\gamma}-D_{\beta}=-ig\left(\sum_{\kappa\cdot\gamma=1,\ 
\kappa\cdot\beta=0}z(\kappa\cdot q)-
\sum_{\kappa'\cdot\beta=1,\ \kappa'\cdot\gamma=0}
z(\kappa'\cdot q)\right).
\]
It is easy to see that there is a one to one correspondence between 
the two summations. For each \(\kappa\) appearing in the first sum we 
define \(\kappa'=\kappa+\beta-\gamma\). Then it is a root 
\(\kappa'=W_{\beta}(\kappa-\gamma)\) and
satisfies \(\kappa'\cdot\beta=1\) and \(\kappa'\cdot\gamma=0\).
Thus we arrive at
\[
D_{\gamma}-D_{\beta}=-ig\sum_{\kappa\cdot\gamma=1,\ \kappa\cdot\beta=0}
\left[z(-\kappa\cdot q)-z((\kappa+\beta-\gamma)\cdot q)\right],
\]
in which the even parity of \(z\) is used.
Now we have
\begin{eqnarray}
	[X,D]_{\beta \gamma}, & = & X_{\beta \gamma}
(D_{\gamma}-D_{\beta})
	\nonumber  \\
	 & = & g^2\sum_{\kappa\cdot\gamma=1,\ \kappa\cdot\beta=0}
	 x((\beta-\gamma)\cdot q)
	 \left[z(-\kappa\cdot q)-z((\kappa+\beta-\gamma)\cdot q)\right]
	 	\label{eq:xdred} \\
	 & = & g^2\sum_{\kappa\cdot\gamma=1,\ \kappa\cdot\beta=0}
	  \left[y((\kappa+\beta-\gamma)\cdot q)x(-\kappa\cdot q)-
	        y(-\kappa\cdot q)
x((\kappa+\beta-\gamma)\cdot q)\right],
\nonumber
\end{eqnarray}
in which the sum rule of the function \(x\), \(y\), and \(z\), 
(\ref{eq:ident1}) is used.

Next we evaluate \([X,Y]_{\beta \gamma}\):
\begin{eqnarray*}
	[X,Y]_{\beta \gamma} & = & \sum_{\kappa\in \Delta,\ 
	\kappa\cdot\beta=1,\ \kappa\cdot\gamma=1}\left(X_{\beta 
	\kappa}Y_{\kappa \gamma}- 
Y_{\beta \kappa}X_{\kappa \gamma}\right) \\
	 & = & -g^2\sum_{\kappa\cdot\beta=1,\ \kappa\cdot\gamma=1}
	 \left[x((\beta-\kappa)\cdot q)y((\kappa-\gamma)\cdot q)-
	 y((\beta-\kappa)\cdot q)x((\kappa-\gamma)\cdot q)\right].
\end{eqnarray*}
By changing the dummy variable from \(\kappa\) to 
\(\kappa'=\gamma-\kappa\), we arrive at
\begin{equation}
	[X,Y]_{\beta \gamma} =-g^2
\sum_{\kappa'\cdot\gamma=1,\ \kappa'\cdot\beta=0}
		 \left[x((\beta-\gamma+\kappa')\cdot q)
y(-\kappa'\cdot q)-
		 y((\beta-\gamma+\kappa')\cdot q)
x(-\kappa'\cdot q)\right],
	\label{eq:xyred}
\end{equation}
which cancels the  previous contribution (\ref{eq:xdred}). It is 
trivial to see that the other terms vanish:
\[
[X_r,D]_{\beta \gamma}=[X,Y_{r}]_{\beta \gamma}
=[X_{r},Y]_{\beta \gamma}=[X_{r},Y_r]_{\beta 
\gamma}=0.
\]
This completes the consistency check for the group 
\(\beta\cdot\gamma=1\).

\subsubsection{(B) \(\beta\cdot\gamma=0\) case}

In this case
\[
[X,D]_{\beta \gamma}=X_{\beta \gamma}(D_{\gamma}-D_{\beta})=0,
\]
since \(X_{\beta \gamma}=0\). The main part also vanishes:
\begin{equation}
	[X,Y]_{\beta \gamma}=\sum_{\kappa\in\Delta}
(X_{\beta \kappa}Y_{\kappa 
	\gamma}-Y_{\beta \kappa}X_{\kappa \gamma})=0.
	\label{eq:xyredb}
\end{equation}
 Suppose there exists a root \(\kappa_1\) such that
\(\beta-\kappa_1\) and \(\kappa_1-\gamma\) are roots (ie, 
\(\kappa_1\cdot\beta=1\) and \(\kappa_1\cdot\gamma=1\)), then 
\(\kappa_2=\beta+\gamma-\kappa_1=W_{\gamma}(\beta-\kappa_1)\) is
a root and satisfies 
\(\beta\cdot\kappa_2=1=\kappa_2\cdot\gamma\).
They always exist  as a pair and their contributions cancel each other
(\(\beta-\kappa_2=\kappa_1-\gamma\) and 
\(\kappa_2-\gamma=\beta-\kappa_1\)):
\begin{eqnarray*}
	&& g^2\left[ x((\beta-\kappa_1)\cdot q)
y((\kappa_1-\gamma)\cdot q)-
	y((\beta-\kappa_1)\cdot q)x((\kappa_1-\gamma)\cdot q)\right.\\
	 & & \quad\quad+ \left.x((\beta-\kappa_2)\cdot q)
	 y((\kappa_2-\gamma)\cdot q)-
	y((\beta-\kappa_2)\cdot q)
	x((\kappa_2-\gamma)\cdot q)\right]=0.
\end{eqnarray*}
It is 
trivial to see that the other terms vanish:
\[
[X_r,D]_{\beta \gamma}=[X,Y_{r}]_{\beta \gamma}
=[X_{r},Y]_{\beta \gamma}=[X_{r},Y_r]_{\beta 
\gamma}=0.
\]
This completes the consistency check for the group 
\(\beta\cdot\gamma=0\).

\subsubsection{(C) \(\beta\cdot\gamma=-1\) case}

This case is a little bit tricky and requires some attention.
In this case again we have
\[
[X,D]_{\beta \gamma}=X_{\beta \gamma}
(D_{\gamma}-D_{\beta})=0=[X_r,D]_{\beta \gamma},
\]
since \(X_{\beta \gamma}=0\) (\((X_r)_{\beta \gamma}=0\)).
But three other terms \([X,Y]\), \([X_r,Y]\), \([X,Y_r]\) have 
non-vanishing contributions.
As for the main term
\[
	[X,Y]_{\beta \gamma}=\sum_{\kappa\in\Delta}
(X_{\beta \kappa}Y_{\kappa 
	\gamma}-Y_{\beta \kappa}X_{\kappa \gamma}),
\]
there is only one intermediate state \(\kappa\). For 
\(\kappa\cdot\beta=1\) and
\(\kappa\cdot\gamma=1\) mean \(\kappa\cdot(\beta+\gamma)=2\), 
implying \(\kappa=\beta+\gamma\). The other two terms \([X_r,Y]\), 
\([X,Y_r]\) have only one intermediate state, too.
Thus we obtain
\begin{eqnarray}
	 [X,Y]_{\beta \gamma}& = & -g^2
\left[x(-\gamma\cdot q)y(\beta\cdot q)-
x(\beta\cdot q)y(-\gamma\cdot q)\right],
	\label{eq:xymin}  \\
\	 [X_r,Y]_{\beta \gamma} & = & -2g^2
\left[x_r(\beta\cdot q)y(-(\beta+\gamma)\cdot q)-
y((\beta+\gamma)\cdot q)x_r(-\gamma\cdot q)\right],
	\label{eq:xrymin}  \\
\	[X,Y_r]_{\beta \gamma} & = & -g^2
\left[x((\beta+\gamma)\cdot 
	q)y_r(-\gamma\cdot q)-
y_r(\beta\cdot q)x(-(\beta+\gamma)\cdot q)\right].
	\label{eq:xyrmin}
\end{eqnarray}
For the functions listed in section two, 
(\ref{eq:functions})  
-- (\ref{eq:functionsell2}) the contribution of 
the above three terms cancel:
\begin{equation}
	[X,Y]_{\beta \gamma}+[X_r,Y]_{\beta \gamma}
	+[X,Y_r]_{\beta \gamma}=0.
	\label{eq:mincancel}
\end{equation}
The nature of the above  condition and its general solutions 
for the elliptic potential case will be discussed elsewhere.

\subsubsection{(D) \(\beta\cdot\gamma=-2\) case}
In this case it is very trivial to see that every term vanishes:
\[
[X,D]_{\beta \gamma}=[X,Y]_{\beta \gamma}=
[X_r,D]_{\beta \gamma}=
[X,Y_{r}]_{\beta \gamma}=[X_{r},Y]_{\beta \gamma}=
[X_{r},Y_r]_{\beta 
\gamma}=0.
\]

Thus the consistency of the Lax pair of  the root type is 
proved for all the four choices of the 
potentials, the rational, trigonometric, hyperbolic and elliptic.

\bigskip
\bigskip
At the end of this section, let us demonstrate that the lowest 
conserved quantity 
is proportional to the Hamiltonian (\ref{eq:hamiltonian}) up to a 
constant:
\begin{equation}
	Tr(L^2)=2I_{Adj}{\cal H}=4h{\cal H},
	\label{eq:consham}
\end{equation}
in which \(I_{Adj}\) is the
{\em second Dynkin index} for the adjoint representation and
$h$ is the Coxeter number, $r+1$, $2(r-1)$ for $A_{r}$ and 
$D_{r}$ and 12, 18 and 30 for $E_{6}$, $E_{7}$ and $E_{8}$.
The second Dynkin index \(I_{\Lambda}\) of {\em any} representation 
\(\Lambda\) is related to the {\em quadratic Casimir invariant} 
\(C_{\Lambda}\) of the representation by
\begin{equation}
	I_{\Lambda}={d_{\Lambda}\over d}C_{\Lambda},
	\label{eq:Dynkin}
\end{equation}
in which \(d_{\Lambda}\) is the dimension of the representation 
\(\Lambda\) and \(d\) is the dimension of the algebra.
Let us fix a root \(\alpha\). Let \(\chi\) be the number of such roots 
\(\beta\) which have unit scalar product with \(\alpha\):
\[
\alpha\cdot\beta=1.
\]
(By the Weyl invariance of the set of the roots, \(\chi\) is the same for 
all roots in \(\Delta\) in the simply-laced theory.)
Then it is easy to see
\begin{eqnarray}
	Tr(E(\alpha)E(\alpha')) & = & \sum_{\beta, \kappa\in\Delta}
	E(\alpha)_{\beta \kappa}E(\alpha')_{\kappa \beta}=
	\sum_{\beta, \kappa\in\Delta}\delta_{\beta-\kappa, \alpha}
	\delta_{\kappa-\beta, \alpha'}
	\nonumber  \\
	 & = & \delta_{\alpha,-\alpha'}\sum_{\beta, \kappa\in\Delta}
	 \delta_{\beta-\kappa, \alpha}
	\nonumber \\
	 & = & \chi\delta_{\alpha,-\alpha'}.
	\label{eq:sumrule1}
\end{eqnarray}
Similarly we have
\begin{equation}
	Tr(E_{d}(\alpha)E_{d}(\alpha'))=\delta_{\alpha,-\alpha'}.
	\label{eq:sumrule2}
\end{equation}
The rest of the trace formulas are trivial:
\[
Tr(E(\alpha))=Tr(E_{d}(\alpha))=Tr(HE(\alpha))=Tr(HE_{d}(\alpha))
=Tr(E(\alpha)E_{d}(\alpha))=0.
\]
We evaluate
\begin{eqnarray*}
	Tr(L^2) & = & Tr[(p\cdot H+X+X_{r})^2] \\
	 & = & Tr[(p\cdot H)^2]+Tr(X^2)+Tr(X_{r}^2)+2Tr(p\cdot HX)+
	 2Tr(p\cdot HX_{r})+2Tr(XX_{r}).
\end{eqnarray*}
The last three terms vanish.
Next we have
\begin{eqnarray*}
	Tr(X^2) & = & -g^2\sum_{\alpha\in\Delta}\sum_{\alpha'\in\Delta}
	x(\alpha\cdot q)x(\alpha'\cdot q)Tr(E(\alpha)E(\alpha'))\\
	 & = & -g^2\chi\sum_{\alpha\in\Delta}x(\alpha\cdot q)x(-\alpha\cdot 
	 q)
\end{eqnarray*}
and
\[
Tr(X_{r}^2)=-4g^2\sum_{\alpha\in\Delta}
x_r(\alpha\cdot q)x_r(-\alpha\cdot q)=
-4g^2\sum_{\alpha\in\Delta}x(\alpha\cdot q)x(-\alpha\cdot q)+const.
\]
In order to evaluate
\[
Tr[(p\cdot H)^2]=\sum_{\beta\in\Delta}(p\cdot\beta)^2,
\]
let us choose \(p\) to be proportional to a fixed root \(\alpha\)
\[
p=\alpha|p|/\sqrt2.
\]
Then we have
\[
Tr(p\cdot\beta)^2={p^2\over2}
\sum_{\beta\in\Delta}(\alpha\cdot\beta)^2
={p^2\over2}(2^2+2^2+\chi+\chi).
\]
In the last expression the first two terms are from
\(\beta=\pm\alpha\) 
and the last two terms are the contributions from 
\(\alpha\cdot\beta=\pm1\).
Thus we arrive at 
\begin{equation}
	Tr(L^2)=(\chi+4)\left(p^2-g^2\sum_{\alpha\in\Delta}
	x(\alpha\cdot q)x(-\alpha\cdot q)\right)
	+const=2(\chi+4){\cal H}+const.
	\label{eq:traceformula}
\end{equation}
By using the  known formula
\begin{equation}
	\chi+4=2h=I_{Adj}
	\label{eq:kaieq}
\end{equation}
we arrive at the result (\ref{eq:consham}).

\section{Lax Pair of the Minimal Type}
\setcounter{equation}{0}

In this section we present a formulation and a proof
of consistency of the minimal type Lax pair.
The proof is valid for all four types of potentials.
This provides a unified framework for all the Calogero-Moser Lax pairs
known  to date.
However, as we will show in a subsequent paper \cite{BCS2} it is 
possible to construct Lax pairs other than the root or the minimal 
types.
Another motivation for this section is to show the close
relationship between the exponents of the algebra and conserved 
quantities. This will be demonstrated explicitly 
in section 4.4 for the \(D_{N}\) models.
The minimal type Lax pairs   have very similar forms to the 
root type Lax pairs.
Their matrix elements are again severely constrained by the
requirement of  Weyl covariance.

\subsection{Minimal Representations}

Let us begin with the definition of the minimal representations
in the theory of Lie algebra representations.
Let \({\mathfrak g}\) be a semi-simple Lie algebra  
(simply  or non-simply laced)
with rank \(r\) and the root system \(\Delta\).
A {\em minimal} representation \(\Lambda\) of \({\mathfrak g}\) 
is an irreducible representation   
such that any weight \(\mu\in\Lambda\)
has scalar products with the roots restricted as follows: 
\begin{equation}
	{2\alpha\cdot\mu\over{\alpha^2}}=0,\pm 1,
\quad \quad \forall\mu\in\Lambda
	\quad \ \mbox{and}\quad \forall\alpha\in\Delta.
	\label{eq:mindef}
\end{equation}
The minimal 
representations have played important roles in various branches of 
physics including conformal field theory \cite{GoOl}.
It is known that the minimal representations are characterised by the 
Coxeter labels and their duals.
For any root \(\alpha\in\Delta\) we define its dual \(\alpha^\vee\) by
\(\alpha^\vee={2\alpha/{\alpha^2}}\). 
Next we introduce the {\em simple roots} 
\(\{\alpha_{1},\ldots,\alpha_r\}\) and the {\em fundamental weights}
\(\{\lambda_{1},\ldots,\lambda_r\}\) as the dual basis to each other:
\begin{equation}
	{2\alpha_{j}\cdot\lambda_{k}\over{\alpha_{j}^2}}=\delta_{jk},\quad
	j,k=1,\ldots,r.
	\label{eq:fundwei}
\end{equation}
The Coxeter labels and their 
duals are
the integers \(n_j\) and   \(n_j^\vee\) appearing in the expansion 
of the highest root 
\(\alpha_0\) in terms of the  simple roots 
\(\{\alpha_{1},\ldots,\alpha_r\}\) and their duals:
\begin{equation}
	\alpha_0=\sum_{j=1}^rn_{j}\alpha_{j},\qquad
	\alpha_0^\vee=\sum_{j=1}^rn_{j}^\vee\alpha_{j}^\vee.
	\label{eq:coxeter}
\end{equation}
A fundamental representation with the highest weight \(\lambda_{j}\)
is {\em minimal} when the corresponding (dual) Coxeter 
label is unity, 
\begin{equation}
	n_{j}=1 \quad  \mbox{or} \quad n_{j}^\vee=1.
	\label{eq:unitcox}
\end{equation}

\bigskip
For the \(A_r\) algebra, all the fundamental representations are 
minimal, \(n_{j}=1,j=1,\ldots,r\). 
The Lax pairs for the \(A_r\) vector and its conjugate representation 
are the first known examples \cite{CalMo,OP1}. 
The Lax pairs for the other fundamental 
representations of \(A_r\) were constructed recently by D'Hoker and 
Phong \cite{DHPh}.
There are three minimal representations of \(D_{r}\). The vector, 
spinor and anti-spinor representations.
The Lax pair for the vector representation has been known 
for many years \cite{OP1}, 
but those for the (anti) spinor representations are 
new \cite{DHPh}.
There are three minimal representations belonging to the simply-laced 
exceptional algebras. The {\bf 27} and \({\bf \overline{27}}\) of \(E_6\)
and {\bf 56} of 
\(E_{7}\). The Lax pairs for these representation are also 
constructed  recently \cite{DHPh}.
The fact that \(E_{8}\) has no minimal representations is 
largely to be blamed for the fact that its integrability has not
been understood earlier. Now the integrability of the  
$E_{8}$ model has been demonstrated above
using the root-type Lax pair.

\bigskip
Among the non-simply laced algebras, 
the vector representations of \(B_r\) and \(C_r\) both have 
unit dual Coxeter 
labels \(n_{V}^\vee=1\). For these, the Lax pairs have been known 
for many years \cite{OP1}. 
D'Hoker and Phong \cite{DHPh} constructed the Lax pair 
for the spinor representation of \(B_r\) which has unit dual Coxeter 
number \(n_{Sp}^\vee=1\). 
The Lax pair of the spinor representation of \(B_r\) 
with two coupling constants can also be obtained easily by 
folding the minimal representation Lax-pair of the spinor and 
anti-spinor representation of \(D_{r+1}\), as we will show presently,
see also \cite{BCS2}.
All the fundamental representations of \(C_r\) have unit dual Coxeter 
labels, \(n_{j}^\vee=1,j=1,\ldots,r\). To the best of our knowledge, 
the Lax pair is known only for the vector representation mentioned 
above.
It is now clear that the Lax pairs for all these representations 
can be easily obtained by folding
the minimal representation Lax pairs of the corresponding 
representations of the \(A_{2r-1}\) algebra.
Lax pairs corresponding to the {\bf 7} dimensional 
representation of \(G_{2}\) and 
{\bf 26} of \(F_{4}\) are also given in \cite{DHPh}.
These representations have unit dual Coxeter labels.
The {\bf 7} dimensional representation of \(G_2\) can be obtained by the
3-fold  reduction of the vector, spinor and anti-spinor representations of
\(D_{4}\). Thus it can also be obtained by folding the minimal
representation  
\(D_{4}\) Lax pairs.
Likewise the {\bf 26} representation of \(F_{4}\) is obtained by 
folding the {\bf 27} and 
 \({\bf \overline{27}}\) of \(E_{6}\).

\subsection{Lax Pair}
Next we construct a  Lax pair in the minimal representation 
\(\Lambda\) 
\begin{equation}
	\Lambda=\{\mu,\nu,\rho, \ldots \},
	\label{eq:minweight}
\end{equation}
of a semi-simple {\em simply-laced} algebra \({\mathfrak g}\)
 with root system \(\Delta\)
of rank \(r\). 
It is invariant under the Weyl group:
\(W_{\alpha}(\mu)\in\Lambda\), \(\forall\mu\in\Lambda\),
 \(\forall\alpha\in\Delta\). It is
known that
\(\Lambda\)  contains no zero weights and that it  consists
of a single Weyl orbit. The Lax pairs have  similar forms to
those of the  root type:
\begin{eqnarray}
	L(q,p) & = & p\cdot H + X ,\nonumber\\
	M(q) & = & D+Y.
	\label{eq:minLaxform}
\end{eqnarray}
Note that, unlike the root type Lax pairs,  \(X_r\) and \(Y_{r}\) 
related with the double roots do not appear.
The matrices \(H\), \(X\) and \(Y\) have the same form as before
\begin{equation}
	X=ig\sum_{\alpha\in\Delta}x(\alpha\cdot q)E(\alpha),\quad
	Y=ig\sum_{\alpha\in\Delta}y(\alpha\cdot q)E(\alpha).
	\label{eq:minXYdef}
\end{equation}
We need only functions $x$, $y$ and $z$ (no $x_{r}$ etc.) and they 
need only satisfy (\ref{eq:ident1}) but not (\ref{eq:mincancel}).
Thus, besides those listed in section two 
(\ref{eq:functions})-(\ref{eq:functionsell6}), 
there are more choices of these functions, for example 
\cite{OP1}:
\begin{equation}
	x(t)={a\over{\sin at}},\quad {a\over{\sinh at}}, 
	\quad {a\over{{\rm sn}(at,k)}},
	\quad a{{\rm cn}(at,k)\over{{\rm sn}(at,k)}},
	\quad a{{\rm dn}(at,k)\over{{\rm sn}(at,k)}}
	\label{eq:morepoten}
\end{equation}
for the trigonometric, hyperbolic and elliptic potentials.
In this section we assume, without loss of generality, that $x$ is an 
odd function while $y$ is even:
\[
x(-t)=-x(t),\quad y(-t)=y(t),\quad y(t)=x^\prime(t).
\]
The difference with the root type Lax pair 
is that their matrix elements are labeled by the 
 weights instead of the roots:
\[
H_{\mu \nu}=\mu\delta_{\mu, \nu},\quad 
	E(\alpha)_{\mu \nu}=\delta_{\mu-\nu,\alpha}.
\]
In the diagonal matrix \(D\) the terms related to the double roots are
dropped:
\begin{equation}
	D_{\mu \nu}= \delta_{\mu, \nu}D_{\mu},\quad
	D_{\mu}=-ig\sum_{\Delta\ni\beta=\mu-\nu}z(\beta\cdot q).
	\label{eq:minD}
\end{equation}
Here the summation is over roots \(\beta\) such that for 
\(\exists\nu\in\Lambda\)
\[
\mu-\nu=\beta\in\Delta.
\]
By multiplying \(\beta\) on both sides, we obtain
\[
\beta^2=2=\beta\cdot\mu-\beta\cdot\nu.
\]
It follows from the assumption of the minimal representation that
the conditions 
\begin{equation}
	\beta\cdot\mu=1\quad \mbox{and} \quad \beta\cdot\nu=-1
	\label{eq:munu}
\end{equation}
must be met in all the terms of \(D_{\mu}\).

As in the case of the root type Lax pair  we rewrite \(D\)
as
\begin{equation}
	D=-ig\sum_{\alpha\in\Delta}z(\alpha\cdot q)K(\alpha),\quad
	K(\alpha)_{\mu \nu}=\delta_{\mu, 
	\nu}\theta(\alpha\cdot\mu),
	\label{eq:minKdef}
\end{equation}
in which \(\theta(t)\) is defined in (\ref{theta}).
Then the Weyl transformation takes the same form as for the root 
type Lax pair:
\begin{equation}
	S(\beta)^{-1}E(\alpha)S(\beta)=E(W_{\beta}(\alpha)),\quad
	S(\beta)^{-1}K(\alpha)S(\beta)=K(W_{\beta}(\alpha)),\quad
	\forall\beta\in\Delta.
	\label{eq:minweylinvXK}
\end{equation}
Here \(S(\beta)\) is the representation matrix of the Weyl 
reflection \(W_{\beta}\) in the weight space \(\Lambda\):
\begin{equation}
	S(\beta)_{\mu \nu}=\delta_{\mu, W_{\beta}(\nu)}.
	\label{eq:minSBform}
\end{equation}
Thus the Weyl covariance of the Lax pair 
(\ref{eq:LMcov}) is guaranteed.

\bigskip
As before we can decompose the Lax equation \(\dot L=[L,M]\) into 
three parts:
\begin{eqnarray}
	{d\over{dt}}X & = & [p\cdot H,Y],
	\label{eq:minqdot}  \\
	{dp\over{dt}}\cdot H & = &
[X,Y]_{\mbox{\footnotesize diagonal part}},
	\label{eq:minpdot}  \\
	0 & = & [X,D+Y]_{\mbox{\footnotesize off-diagonal
part}}.
	\label{eq:minoffd}
\end{eqnarray}
The first equation (\ref{eq:minqdot}) is equivalent to the 
first half of the canonical 
equations of motion (\ref{eq:caneq}) \(\dot q=p\) and it can 
be shown in the same 
way as for the root type Lax pairs.
Next we show that the second equation (\ref{eq:minpdot}) is equivalent 
to the second half of the canonical equations of motion
(\ref{eq:caneq}). The proof is valid for all four types of potentials.
\begin{eqnarray*}
	[X,Y]_{\mu \mu} & = & \sum_{\nu\in\Lambda}
	(X_{\mu\nu}Y_{\nu\mu}-Y_{\mu\nu}X_{\nu\mu})  \\
	 & = & -g^2\sum_{\nu\in\Lambda}\sum_{\alpha\in\Delta}
\sum_{\beta\in\Delta}
	 x(\alpha\cdot q)y(\beta\cdot 
	 q)\left(\delta_{\mu-\nu,\alpha}\delta_{\nu-\mu,\beta}-
	 \delta_{\mu-\nu,\beta}\delta_{\nu-\mu,\alpha}\right)\\
	 & = & -2g^2\sum_{\alpha\in\Delta,\ \alpha\cdot\mu=1}
	 x(\alpha\cdot q)y(\alpha\cdot q),
\end{eqnarray*}
in which the parity of functions \(x\) (odd) and \(y\) (even) is used.
Thus we obtain from (\ref{eq:minpdot}):
\begin{equation}
	\dot p\cdot\mu= -2g^2
\sum_{\alpha\in\Delta,\ \alpha\cdot\mu=1}
		 x(\alpha\cdot q)y(\alpha\cdot q).
	\label{eq:mineq}
\end{equation}

On the other hand, by multiplying \(\mu\) on both sides of the 
second Hamiltonian equation (\ref{eq:caneq}), we obtain:
\[
	\dot p\cdot\mu=-g^2\sum_{\alpha\in\Delta}
x(\alpha\cdot q)y(\alpha\cdot 
	q)\alpha\cdot\mu.
\]
By assumption of the minimal representation 
\(\alpha\cdot\mu\) takes 
value 0 or \(\pm1\) and only the latter contribute:
\begin{eqnarray}
	 \dot p\cdot\mu& = & -g^2\left(\sum_{\alpha\in\Delta,\ 
	 \alpha\cdot\mu=1}x(\alpha\cdot q)y(\alpha\cdot q)-
	 \sum_{\alpha\in\Delta,\ 
	 \alpha\cdot\mu=-1}x(\alpha\cdot q)y(\alpha\cdot q)\right)
	\nonumber  \\
	 & = & -2g^2\sum_{\alpha\in\Delta,\ \alpha\cdot\mu=1}
		 x(\alpha\cdot q)y(\alpha\cdot q),
	\label{eq:mineqfin}
\end{eqnarray}
in which the parity of the functions is used. Thus the equivalence to 
the canonical equations of motion is proved.

Next we show (\ref{eq:minoffd}), or the consistency of the Lax pair.
First we have (\(\mu\neq\nu\))
\begin{eqnarray}
	 [X,D]_{\mu \nu} 
	 & = & X_{\mu \nu}(D_{\nu}-D_{\mu})
	\nonumber \\
	 & = & g^2\sum_{\alpha\in\Delta}
x(\alpha\cdot q)E(\alpha)_{\mu\nu}\left(
	 \sum_{\beta\in\Delta,\ \beta\cdot\nu=1}
z(\beta\cdot q)-
	 \sum_{\beta\in\Delta,\ \beta\cdot\mu=1}
z(\beta\cdot q)\right),
	\label{eq:minxd}
\end{eqnarray}
which vanishes if \(\mu-\nu\) is not a root.
Likewise the main part
\begin{equation}
	[X,Y]_{\mu \nu}=-g^2
\sum_{\rho\in\Lambda,\ \mu-\rho\in\Delta,\ 
	\rho-\nu\in\Delta}\left[x((\mu-\rho)\cdot q)
y((\rho-\nu)\cdot q)-
	y((\mu-\rho)\cdot q)x((\rho-\nu)\cdot q)\right]
	\label{eq:minxy}
\end{equation}
vanishes if \(\mu-\nu\) is not a root. 
This can be shown in a similar way to (\ref{eq:xyredb}).
If \(\mu-\nu\in\Delta\) we can use the sum rule of the functions 
(\ref{eq:ident1}) to obtain
\begin{eqnarray}
	 [X,Y]_{\mu \nu}& = & -g^2
\sum_{\rho\in\Lambda}x((\mu-\nu)\cdot q)
	 \left[z((\rho-\nu)\cdot q)-z((\mu-\rho)\cdot q)\right]
	 	\nonumber \\
	 & = & -g^2x((\mu-\nu)\cdot q)\sum_{\rho\in\Lambda}
	 \left[z((\nu-\rho)\cdot q)-z((\mu-\rho)\cdot q)\right]
	\nonumber \\
	 & = & -g^2\sum_{\alpha\in\Delta}
x(\alpha\cdot q)E(\alpha)_{\mu \nu}
	 \left(\sum_{\beta\in\Delta,\ \beta\cdot\nu=1}
z(\beta\cdot q)-
	 \sum_{\beta\in\Delta,\ \beta\cdot\mu=1}
z(\beta\cdot q)\right).
	\label{eq:minxydef}
\end{eqnarray}
This cancels the above expression (\ref{eq:minxd}) and the 
consistency is proved.

\bigskip
At the end of this subsection, let us remark that the relationship 
between  the
lowest conserved quantity and the Hamiltonian (\ref{eq:hamiltonian})
takes the same form as in the root type Lax formulation (\ref{eq:consham}):
\begin{equation}
	Tr(L^2)=2I_{\Lambda}{\cal H},
	\label{eq:minconsham}
\end{equation}
in which $I_{\Lambda}$ is as before the second Dynkin index (\ref{eq:Dynkin})
of the representation \(\Lambda\).
The derivation is similar and rather easier than the case of the root 
type Lax pairs and therefore it will not be repeated.
One only has to note the following relation
\begin{equation}
	 \chi_{\Lambda}=I_{\Lambda}
	\label{eq:kailameq},
\end{equation}
in which \(\chi_{\Lambda}\) is the number of such weights \(\mu \in
\Lambda\)
 that they have a unit scalar product with a fixed root \(\alpha\):
\[
\alpha\cdot\mu=1.
\]

\bigskip
In the rest of this section we show that the minimal representations
give the known examples of Calogero-Moser Lax pairs by choosing some 
typical cases. 
We also remark that the correspondence between the conserved 
quantities and the exponents of the algebra can be seen 
most clearly in the Lax pairs of  the spinor
and anti-spinor representations of the \(D_{N}\) theory.

\subsection{$A_{N-1}$ Vector Representation}
We introduce an $N$ dimensional orthonormal basis of \({\mathbb R}^N\)
\begin{equation}
	e_{j}\cdot e_{k}=\delta_{j,k},\quad j,k=1,\ldots,N.
	\label{eq:nbasis}
\end{equation}
Then the sets of roots and vector weights
\footnote{To be more precise, the weight is \(e_{j}-\mu_{0}\), 
\(\mu_{0}=(e_{1}+\cdots+e_{N})/N\). We assume, without loss of
generality, that the
system is in the center of mass frame: \(p_1+\cdots+p_N=0\).
Then the  \(\mu_0\) part
does not contribute to the Lax equation or to the Hamiltonian
(\ref{eq:minconsham}).} are:
\begin{eqnarray}
	\Delta&=&\{e_{j}-e_{k}:\quad j,k=1,\ldots,N\}, \quad 
	\nonumber\\
	\Lambda&=&\{e_j:\quad j=1,\ldots,N\}.
	\label{eq:ANdel}
\end{eqnarray}
The Weyl group is represented simply by a permutation of $N$ elements:
\begin{equation}
	S(e_j-e_k)=P(j,k),
	\label{eq:Nperm}
\end{equation}
in which $P(j,k)$ is the $N\times N$ matrix for permuting $j$ and $k$.
The matrices \(E\) and \(K\) are:
\begin{equation}
	E(e_{j}-e_k)_{lm}=\delta_{j,l}\delta_{k,m},\quad
	K(e_{j}-e_k)_{lm}=\delta_{l,m}\left(\delta_{j,l}+\delta_{k,m}\right).
	\label{eq:ANXKvec}
\end{equation}
In this basis the Lax pair takes the well-known form \cite{OP1}:
\begin{eqnarray}
	L_{jk}&=&p_{j}\delta_{j,k}+ig(1-\delta_{j,k})\,x(q^j-q^k),\quad 
	M_{jk}=D_{j}\delta_{j,k}+ig(1-\delta_{j,k})\,y(q^j-q^k),\nonumber\\
	\quad D_{j}&=&-ig\sum_{k\neq j}z(q^j-q^k).
	\label{eq:ANLM}
\end{eqnarray}

\subsection{$D_{N}$}
The set of roots in the above orthonormal basis (\ref{eq:nbasis}) is
\begin{equation}
	\Delta=\{e_{j}-e_{k},\quad \pm(e_j+e_k) :\quad 
j,k=1,\ldots,N \}. 
	\label{eq:DNdel}
\end{equation}

\subsubsection{Vector Representation}
In the vector representation $\Lambda$ has $2N$ dimensions
\[
\Lambda=\{e_j, -e_j: \quad j=1,\ldots, N\}.
\]
The Weyl group consists of  permutations of $N$ elements 
and a sign change:
\begin{equation}
	S(e_j-e_k)=1\otimes P(j,k),\quad
	S(\pm(e_j+e_k))=P(+,-)\otimes P(j,k),
	\label{eq:DNperm}
\end{equation}
in the first set the permutation acts on both positive and negative 
weights.
In the second set the positive weights and the negative weights 
are permuted together with an exchange of \(j\) and \(k\).
The matrices \(E\) are those given in the literature:
\begin{equation}
	E(e_{j}-e_k)_{lm}=\delta_{l, j}\delta_{m, k}+
	\delta_{l, k+N}\delta_{m, j+N},
\end{equation}
\begin{equation}
	E(e_{j}+e_k)_{lm}=\delta_{l, j}\delta_{m, k+N}+
	\delta_{l, k}\delta_{m, j+N},\quad E(-e_j-e_k)=E(e_j+e_k)^T,
	\label{DNXvec}
\end{equation}	
\begin{equation}
	K(\pm e_{j}\pm e_k)_{lm}=
	\delta_{l,m}\left(\delta_{j,l}+\delta_{k,m}+
\delta_{l, j+N}+\delta_{m, k+N}\right).
	\label{eq:DNKvec}
\end{equation}
This gives the well known Lax pair in the block notation \cite{OP1}:
\begin{equation}
L=\left(\begin{array}{cc}
A_{1}& B_{1} \\ -B_{1} & -A_{1} 
\end{array}\right),
\quad
M=\left(\begin{array}{cc}
A_{2}& B_{2} \\ B_{2} & A_{2} 
\end{array}\right),
 \label{eq:DNVLM}    
\end{equation}
in which
\begin{eqnarray}
(A_{1})_{jk}&=&p_j\delta_{j,k}+ig(1-\delta_{j,k})\,x(q^j-q^k),\quad 
(B_{1})_{jk}=ig(1-\delta_{j,k})\,x(q^j+q^k),\nonumber\\
(A_{2})_{jk}&=&D_j\delta_{j,k}+ig(1-\delta_{j,k})\,y(q^j-q^k),\quad 
(B_{2})_{jk}=ig(1-\delta_{j,k})\,y(q^j+q^k),\nonumber\\
D_{j}&=&-ig\sum_{k\neq j}\left[z(q^j-q^k)+z(q^j+q^k)\right].
\label{eq:DNVdet}
\end{eqnarray}

\subsubsection{Spinor plus Anti-Spinor Representations}
Each of the spinor and anti-spinor representations has \(2^{N-1}\) 
dimensions.
Instead of writing down the matrix elements of \(L\) and \(M\) in 
each of these 
representations, we choose to express the Lax pair in 
a more conventional form using the Pauli matrices acting on 
the tensor product of two component spinors,
\({\mathbb C}^2\otimes\cdots\otimes{\mathbb C}^2\) (\(N\) times).
It is a reducible representation 
of spinor\(\oplus\)anti-spinor representations. The sets of weights
$\Lambda$ has $2^N$ dimensions:
\[
\Lambda=\{{1\over2}\sum_{j=1}^N\epsilon_{j}e_j:
\quad \epsilon_{j}=\pm1,\quad j=1,\ldots,N\}.
\]
The matrices \(E\) and \(K\)  are:
\begin{equation}
	E(e_j-e_k)=\sigma_+^{(j)}\sigma_-^{(k)},\quad
	E(e_j+e_k)=\sigma_+^{(j)}\sigma_+^{(k)},\quad
	E(-e_j-e_k)=\sigma_-^{(j)}\sigma_-^{(k)},
	\label{eq:DNXsp}
\end{equation}
\begin{equation}
	K(e_j-e_k)={1\over4}(1-\sigma_z^{(j)}\sigma_z^{(k)}),\quad
	K(\pm(e_j+e_k))={1\over4}(1+\sigma_z^{(j)}\sigma_z^{(k)}).
	\label{eq:DNKsp}
\end{equation}
In the above expressions \(\sigma_x\), \(\sigma_z\) and 
\(\sigma_\pm\) 
 are Pauli sigma matrices:
\[
\sigma_x=\left(
\begin{array}{cc}
	 0& 1  \\
	1 & 0
\end{array}\right),\quad 
 \sigma_z=\left(\begin{array}{cc}
	 1& 0  \\
	0 & -1
\end{array}\right),\quad 
 \sigma_+=\left(\begin{array}{cc}
	 0& 1  \\
	0 & 0
\end{array}\right),\quad 
\sigma_-=\left(\begin{array}{cc}
	 0& 0  \\
	1 & 0
\end{array}\right).
\]
The superscripts on the Pauli matrices denote on which space to act.
The Weyl group consists of  permutations of $N$ elements and an 
exchange of plus and 
minus signs:
\begin{equation}
	S(e_j-e_k)=P(j,k),\quad
	S(\pm(e_j+e_k))=P(+,-)\otimes \sigma_x^{(j)}\sigma_x^{(k)}.
	\label{eq:DNweylsp}
\end{equation}
The Lax pair is expressed simply as
\begin{eqnarray}
	L_{D_{N}} & = & {1\over2}\sum_{j=1}^Np_{j}\sigma_{z}^{(j)}
	                +ig\sum_{j<k}x(q^j-q^k)
					(\sigma_{+}^{(j)}\sigma_{-}^{(k)}-
					\sigma_{+}^{(k)}\sigma_{-}^{(j)})\nonumber\\
	 &  & \quad \quad \quad \quad 
\ \ \ \,    +ig\sum_{j<k}x(q^j+q^k)
					(\sigma_{+}^{(j)}\sigma_{+}^{(k)}-
					\sigma_{-}^{(k)}\sigma_{-}^{(j)}),
	\nonumber  \\
	 M_{D_{N}}& = & -ig\sum_{j<k}z(q^j-q^k){1\over2}
	 (1-\sigma_{z}^{(j)}\sigma_{z}^{(k)})
	 -ig\sum_{j<k}z(q^j+q^k){1\over2}
	 (1+\sigma_{z}^{(j)}\sigma_{z}^{(k)})
	  \label{eq:DNspLM}\\
	 &  & +ig\sum_{j<k}y(q^j-q^k)(\sigma_{+}^{(j)}\sigma_{-}^{(k)}+
	 \sigma_{+}^{(k)}\sigma_{-}^{(j)})
	 +ig\sum_{j<k}y(q^j+q^k)(\sigma_{+}^{(j)}\sigma_{+}^{(k)}+
	 \sigma_{-}^{(j)}\sigma_{-}^{(k)}).
	\nonumber
\end{eqnarray}
The spinor and anti-spinor representations are characterised by the
eigenvalues (\(\pm1\)) of the following matrix \(\Gamma\) 
($\gamma_{2N+1}$, 
the analogue of $\gamma_5$ in four dimensions):
\begin{equation}
	\Gamma=\prod_{j=1}^N\sigma_{z}^{(j)},\quad \Gamma^2=1,\quad
	\Gamma^\dagger=\Gamma.
	\label{eq:gamma5}
\end{equation}
It commutes with all the matrices appearing in \(L\) and \(M\).
Thus by using projectors 
\begin{equation}
	P_{\pm}={1\over2}(1\pm\Gamma), \quad P_{\pm}^\dagger=
P_{\pm}^2=P_{\pm},
	\label{eq:proj}
\end{equation}
the Lax pairs in the spinor and anti-spinor representations are
obtained as
\[
LP_{\pm} \quad \mbox{and} \quad MP_{\pm}.
\]
Thus we have two sets of conserved quantities for the 
\(D_{N}\)  Calogero-Moser model, 
\begin{equation}
	Tr(L^kP_{+}) \quad  \mbox{and} \quad Tr(L^kP_{-}), \quad k=1,\ldots,
	\label{eq:DN cons}
\end{equation}
derived from the spinor and anti-spinor representations, respectively.
However, for most values of \(k\), except for \(N\), they give the same 
conserved quantities
since the difference vanishes:
\begin{equation}
		Tr(L^k\Gamma)=0,\quad  \mbox{except for }\quad k=N.
		\label{eq:nulls}
\end{equation}
Among the conserved quantities \(Tr(L^k)\) of Toda theories and 
Calogero-Moser models
based an a Lie algebra \({\mathfrak g}\),
the independent ones are known to occur at \(k\) equal 
to the {\em exponent} of  
\({\mathfrak g}\) plus 1. (At the other values of \(k\), \(Tr(L^k)\) 
either vanishes or is a  polynomial of the lower order conserved 
quantities.)
For every Lie algebra 1 is always an exponent.
This corresponds to the universal fact that the lowest conserved
quantity \(Tr(L^2)\) is (proportional to) the Hamiltonian 
(\ref{eq:consham}), (\ref{eq:minconsham}).
For \(D_{N}\) the exponents are \((1,3,\ldots,2N-3,N-1)\) and 
in the present case the exponents
\((1,3,\ldots,2N-3)\) correspond to the conserved quantities
\(Tr(L^{2k})\), \((k=1,2,\ldots,N-1)\) obtained in the 
spinor or anti-spinor representations; 
the exponent \(N-1\) corresponds to
the extra conserved quantity derived above
\begin{equation}
	Tr(L^N\Gamma).
	\label{eq:extra}
\end{equation}

\subsection{\(E_{6}\) and \(E_{7}\)}

The {\bf 27} and \({\bf \overline{27}}\) dimensional representations of
\(E_{6}\) are minimal. In both cases \(\Lambda\) is decomposed into 
\({\bf 1}+{\bf 10}+{\bf 16}\) or the singlet  plus the vector plus the
spinor representations of
\(D_{5}\).

The minimal {\bf 56}  dimensional representations of \(E_{7}\)
is decomposed into \({\bf 12}+{\bf 32}+{\bf 12}\).  That is the sum of two
vector  representations and
a spinor representation of \(D_{6}\). 

In both cases the structure of the Lax pair in each sector is 
described as above.
We have not yet found a more succinct way of representing their Lax 
pair than the general form of the minimal 
type (\ref{eq:minLaxform}).

\bigskip
Let us give some simple examples of the Lax pairs of the non-simply 
laced algebras obtained by reduction (folding) of the minimal 
representation ones for the simply-laced algebras.

\subsection{Some \(B_{N}\) Lax Pairs by Reduction}
It is possible to obtain \(2N+2\) dimensional representation of
the \(B_{N}\) Lax pair in the vector representation. One only has to 
impose restrictions on the dynamical variables in the Lax pair of 
\(D_{N+1}\) in the vector representation:
\begin{equation}
	q^{N+1}=p_{N+1}=0.
	\label{eq:restr}
\end{equation}
This representation can be easily reduced to the well known one \cite{OP1} 
in \(2N+1\) dimensions with the coupling constant of the short roots 
given by \(g_{1}=\sqrt2 g\).

It is more interesting to derive the Lax pair of the spinor representation
of \(B_{N}\) from the spinor\(\oplus\)anti-spinor representation of 
\(D_{N+1}\) given above.
Together with the restriction of the dynamical variables as above 
(\ref{eq:restr}) one can also impose \(\sigma^{(N+1)}\to1\) 
to obtain the \(2^N\) dimensional representation:
\begin{equation}
	L_{B_N}=L_{D_N}+ig_{1}\sum_{j=1}^Nx(q^j)
(\sigma_+^{(j)}-\sigma_-^{(j)}),\quad
	M_{B_N}=M_{D_N}+ig_{1}\sum_{j=1}^Ny(q^j)
(\sigma_+^{(j)}+\sigma_-^{(j)}).
	\label{eq:BXSred}
\end{equation}
It is elementary to verify that the coupling constant of the short roots
\(g_{1}\) can be independent of \(g\). (The reduction itself gives the
relation \(g_{1}= g\).) 

\subsection{Vector Representation of \(C_{N}\) Lax Pair by Reduction}
The  simplest example of a Lax pair derived by reduction is 
that of the vector representation of \(C_{N}\).
Starting from the \(A_{2N-1}\) vector representation   
Lax pair and imposing  restrictions on the dynamical variables:
\begin{equation}
	q^{2N+1-j}=-q^j,\quad p_{2N+1-j}=-p_j,\quad j=1,\ldots,N,
	\label{eq:CNred}
\end{equation}
we obtain the well-known form of the Lax pair \cite{OP1}:
\begin{equation}
L=\left(\begin{array}{cc}
A_{1}& B_{1} \\ -B_{1} & -A_{1} 
\end{array}\right),
\quad
M=\left(\begin{array}{cc}
A_{2}& B_{2} \\ B_{2} & A_{2} 
\end{array}\right),
 \label{eq:CNVLM}    
\end{equation}
in which
\begin{eqnarray}
(A_{1})_{jk}&=&p_j\delta_{j,k}+ig(1-\delta_{j,k})\,x(q^j-q^k),
\ \ (B_{1})_{j,k}=ig(1-\delta_{j,k})\,x(q^j+q^k)
 + ig_{4}\,x(2q^j)\delta_{j,k},\nonumber\\
(A_{2})_{jk}&=&D_j\delta_{j,k}+ig(1-\delta_{j,k})\,y(q^j-q^k),\ \ 
(B_{2})_{jk}=ig(1-\delta_{j,k})\,y(q^j+q^k)+ig_{4}\,
y(2q^j)\delta_{j,k},\nonumber\\
D_{j}&=&-ig\sum_{k\neq 
j}\left[z(q^j-q^k)+z(q^j+q^k)\right]-ig_4\,z(2q^j).
\label{eq:CNVdet}
\end{eqnarray}
In this case the representation space has dimension \(2N\) and
it is elementary to verify that the coupling constant of the long roots
\(g_{4}\) can be independent of \(g\). 

Other types of reductions and the formulation of minimal representation
in non-simply laced algebra will be discussed elsewhere \cite{BCS2}.

\section{Summary and Comments}
\setcounter{equation}{0}
A simple and {\em universal Lax pair} for the Calogero-Moser 
models based on any semi-simple Lie algebras, including \(E_{8}\), is
presented. It is based on the root system and Weyl invariance only,
suggesting  the possibility of generalising Calogero-Moser models to a
wider  class of root systems beyond those associated with Lie algebras.
The key idea is the representation of the Weyl reflections on the 
set of roots itself for the root type Lax pair.
Thus it is applicable, in principle, to all Calogero-Moser models.
The proof of the consistency of the Lax pair
is elementary and it has been checked 
for all the four types of interaction potentials.
As for the representation of the Weyl reflections, 
the root type Lax pair is conceptually better than 
the adjoint representation which 
consists of the set of roots and the zero weights corresponding to the
Cartan subalgebra.
If the zero weights were included, the representation matrices of 
Weyl reflections could not be unique on them.
This does not mean, however, that the Lax pairs in the adjoint
representations do not  exist.
We will report some examples of Calogero-Moser Lax pairs in
adjoint representations and symmetric tensor representations
in a subsequent paper \cite{BCS2}.

\bigskip
Another type of Calogero-Moser Lax pair, called  minimal 
type, is introduced.
The minimal types provide a unified description of all
Calogero-Moser Lax pairs known to date and reveals some new ones.
Lax pairs belonging to the minimal type of
non-simply laced theories are related to those of the simply-laced 
theories by reduction.
The spinor\(\oplus\)anti-spinor representations of \(D_{N}\) models
are discussed in some detail in connection with the an 
alternative representation of the conserved 
quantities and with the reduction to the \(B_{N}\) spinor
representation.

\bigskip
Since the non-Lie algebraic aspects of the Calogero-Moser models
are highlighted, it would be interesting to see if these models
could be obtained by reduction of self-dual Yang-Mills equations
related with some Lie algebras \cite{folk}.

\begin{center}
{\bf ACKNOWLEDGMENTS}
\end{center}
We thank H.\,W.\, Braden and D.\, Olive for useful discussion.
This work was supported by the Anglo-Japanese Collaboration Project
of the Royal Society and the Japan Society for the Promotion of Science.
A.\,J.\,B is supported by the Japan Society for the Promotion of Science 
and the National Science Foundation under grant no. 9703595.

\end{document}